# IMPROVING COMPUTER-MEDIATED SYNCHRONOUS COMMUNICATION OF DOCTORS IN RURAL COMMUNITIES THROUGH CLOUD COMPUTING: A CASE STUDY OF RURAL HOSPITALS IN SOUTH AFRICA

Alfred Coleman [1], Marlien E Herselman [2] and Mary Coleman [3]

[1]School of Computing, University of South Africa
colemana@iburst.co.za
[2] Meraka Institute, CSIR South Africa
mherselman@csir.co.za
[3] Medical University of South Africa
marycoleman@webmail.co.za

*ABSTRACT*

*This paper investigated how doctors in remote rural hospitals in South Africa use computer-mediated tool to communicate with experienced and specialist doctors for professional advice to improve on their clinical practices. A case study approach was used. Ten doctors were purposively selected from ten hospitals in the North West Province. Data was collected using semi-structured open ended interview questions. The interviewees were asked to tell in their own words the average number of patients served per week, processes used in consultation with other doctors, communication practices using computer-mediated tool, transmission speed of the computer-mediated tool and satisfaction in using the computer-mediated communication tool. The findings revealed that an average of 15 consultations per doctor to a specialist doctor per week was done through face to face or through telephone conversation instead of using a computer-mediated tool. Participants cited reasons for not using computer-mediated tool for communication due to slow transmission speed of the Internet and regular down turn of the Internet connectivity, constant electricity power outages and lack of e-health application software to support real time computer-mediated communication. The results led to the recommendation of a hybrid cloud computing architecture for improving communication between doctors in hospitals.*

## 1. INTRODUCTION

The delivery of healthcare services in Africa is severely affected by massive shortages of doctors, inadequate deployment of Information and Communication Technology (ICT) infrastructure and unattractive incentives to doctors from universities who are deployed to remote rural hospitals in Africa. These doctors are often deployed to remote locations where access to knowledge sharing facilities to improve their work performance becomes impossible. Computer-mediated communication systems (e-mail, telemedicine system and video conferencing) which are ICT components have become the foremost tools to bridge the gap between doctors in remote locations and specialist doctors in urban areas [1] .These tools provide support to doctors in remote rural areas and they are not just mere mechanism for exchange of knowledge but a mechanism for creating knowledge repository and a mechanism for accessing the knowledge repository [2]. Sharing knowledge and putting the shared knowledge into a repository source is an important start in knowledge-sharing and a basis for organizational memory [3; 4]. However, a number of factors influence the efficiency of knowledge sharing which include the characteristics of the knowledge being shared and those of the channel being used [5].

Computer-mediated communication system is a channel that can support and promote such knowledge sharing among health professionals [1]. Despite the compelling advantages of computer-mediated tools in healthcare, doctors who are deployed to remote regions in South Africa are unable to use neither real time consultation nor asynchronous communication through computer-mediated tool to communicate with their counterparts who are specialist doctors at provincial or teaching hospitals in South Africa for professional advice.

This paper investigated how doctors in remote community hospitals in South Africa use computer-mediated tools to communicate with experienced and specialist doctors for professional advice and based on the findings proposed a service oriented architectural framework (cloud computing) to promote asynchronous and real time communication to improve performance of doctors who are in remote rural areas.

The proceeding sections of this paper are presented as follows: literature and theoretical framework; methods, results, discussion and proposed architectural framework (cloud computing) and finally conclusion.

## 2. LITERATURE AND THEORETICAL FRAMEWORK

Synchronous communication is direct communication where the communicators are present at the same time. This includes, but is not limited to, a telephone conversation, a company board meeting, a chat room event and instant messaging. Synchronicity exists among individuals when they exhibit a shared pattern of coordinated synchronous behaviour with a common focus [6]. Asynchronous communication on the other hand, does not require all parties involved in the communication to be present at the same time. Asynchronous computer- mediated communication is limited to communication where a network infrastructure capability of sustaining real time media connection is not cost-effective [7].Asynchronous remote communication systems can be divided into the following: message, storage and discussion-centric systems. Message-centric systems are those which function like e-mails [8]. It enables doctors to send questions to specialist doctors and receive replies. They are easy to use but lack content management features that are useful to establish a community of communicators. Storage-centric systems are like Web or message-based picture archive communication systems that are often used in teleradiology [9]. It adds basic search and storage capabilities. Discussion-centric systems implement the functionality of a typical Web-based bulletin-board system (incorporating messaging, discussion, and image storage facilities) which allows for two way communication. This paper, therefore, positions asynchronous computer-mediated communication between storage and discussion-centric system.

Recent research on communication has focused on Media Synchronicity Theory (MST) which is propounded by[10]. MST is defined as the extent to which the capabilities of a communication medium enable individuals to achieve synchronicity[11] .Synchronicity is a state in which individuals work together at the same time with common focus [10]. MST marches media capabilities to the relevant communication processes and can therefore be used to study communication performance [10].MST suggests that supporting media should fit two fundamental communication processes. These are conveyance processes which lead to individual understanding and convergence processes which lead to shared understanding. According to [12] communication media should not only support the transmission of information but also the manifestation of meaning. MST has two sub processes necessary for conveyance and convergence of communication processes. These are information transmission and information processing. The diagram below illustrates these processes.

Source: [10]

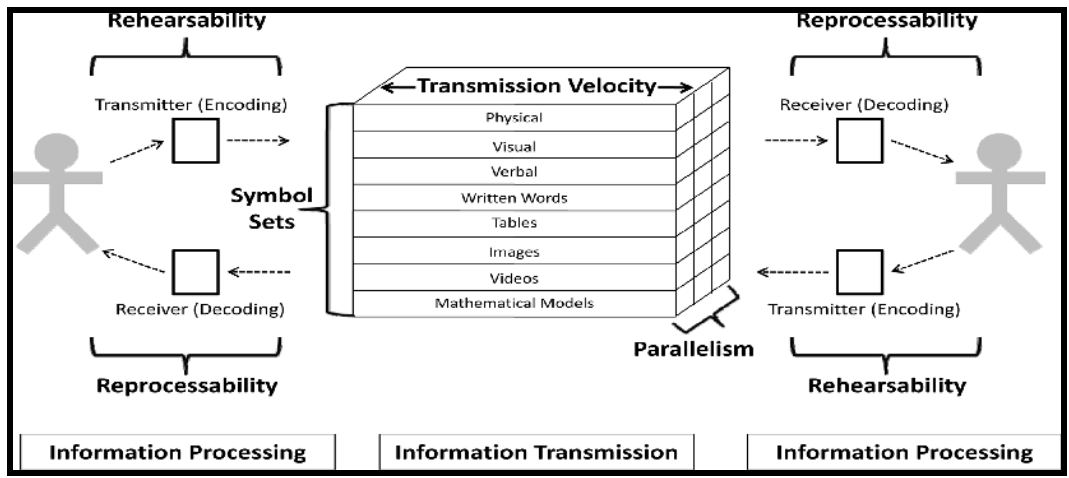

Figure1: Communication processes

The process starts with the source, the sender of the message, who uses a transmitter to encode the message into a signal. The message is sent over a communication channel (medium) to a receiver, which is used by the recipient (destination) to decode the signal back to the original message. Dennis and others) introduce five media capabilities that influence information processing and transmission [10]. These are rehearsability, reprocessability, transmission velocity, parallelism, and symbols set.

Rehearsability is the extent to which the media enables the sender to rehearse or fine tune a message before sending it. On the other side of the transmission process is reprocessability which enables a message to be re-examined or processed again. Transmission velocity refers to the speed at which a medium can deliver a message to intended recipients. The forth media capability is parallelism, which enables a number of simultaneous transmissions from multiple senders to take place effectively. This influences the quantity of information that can be transmitted within a certain time. Finally the media capability of symbols set refers to the number of ways in which a medium allows information to be coded for communication. These five capabilities influence synchronicity for conveyance and convergence processes.

## 2.1. Cloud computing

Cloud computing provides IT capabilities to organizations via a network. The notion of cloud computing refers to a technology infrastructure model that enables several types of computing tasks to be performed over a network[13]. A cloud computing infrastructure is composed of several types of hardware components (e.g. servers, storage systems and network components), software application components (programs, services and protocols) and information [14]. The network can be a local area network or wide area network like the Internet and the network depends on the type of cloud, a private cloud, public cloud or a hybrid cloud computing services.

A private cloud infrastructure is created and deployed only for a single organization and can be managed by the same organization or by a third party while a public cloud enables general access to all organizations. A hybrid cloud combines the features of public and private clouds and delivers a single cloud solution.

Cloud computing is driven by economies of scale in which there is a pool of virtualized and dynamically-scalable computing power, providing storage, platforms and services on demand to organizations [14]. It represents a new paradigm for organizations' IT operations where pay-as-you-consume business model is used. Again the benefits of cloud computing to organizations include cost reduction, high automation, scalability and efficient purchasing management.

When health organization considers moving its services into the clouds, the organization needs a health care cloud computing strategic plan model [15]. Kuo proposes a strategic plan model which comprises four stages (identification, evaluation, action and follow up) for health organization [15]. Identification means the health care organization analysis the current status of the health organizations process and identifies the fundamental objective of services improvement. Evaluation which is the second stage means to evaluate the opportunities and challenges of adopting cloud computing. After the evolution stage is the action stage where the organization determines whether to adopt the service or not. If the organization adopts cloud computing it draws up an action plan. The final stage is the follow up stage which deploys cloud computing infrastructure and develop a follow up- plan.

## 3. Methods

The study was carried out in the North West Province of South Africa. Ten community hospitals (Taung, Ganyesa, Revilon, Bloemhof, Klerksdorp, Rustenburg, Christiana, Boitumelong, Empilisweni and Classic House Hospitals) in the North West Province of South Africa were purposefully selected. These hospitals were selected based on their geographical locations which span across the entire province and form part of the government owned institutions in South Africa.

The participants for the study were drawn from the entire population of doctors in the ten hospitals. In describing population [16] indicate that it is the aggregate of cases having a common and designated criterion that is accessible as subjects for a study. A purposive sampling technique was used in selecting the participants. A doctor from each of these hospitals was selected. The participants were selected by their professions which was relevant to the study. Ten doctors volunteered to participate in the study.

Data was collected using semi-structured open ended interviews. The interviewees represented different roles ranging from specialist doctors to general practitioners. The interviewees were asked to tell in their own words the average number of patients served per week, processes used in consultation with other doctors, communication practices using computer-mediated tool, transmission speed of the computer-mediated tool (CMT) and satisfaction using the computer-mediated communication tool. The interview lasted for one and a half hours with each interviewee and was audio-recorded and transcribed by the researcher.

Integrity of data entry from the study was checked by another researcher. Transcripts were coded using [17] method of case study analysis techniques. After the initial coding, broad categories were identified by searching for patterns in the participants' responses. The categories were ICT infrastructure availability, doctor's workload, methods of consultation among doctors, frequency of consultation, task type, speed of transmission of task using computer-mediated tool and satisfaction using a computer-mediated tool.

## 4. RESULTS

The results are presented under the categories of ICT infrastructure availability, doctor's workload, methods of consultation among doctors and frequency of consultation, task type and speed of transmission of task using computer-mediated tool and satisfaction using computer-mediated tool.

### 4.1. ICT infrastructure availability

The findings revealed that there are ICT equipment like computers, fax machines, telephones, scanners and Internet connections in the hospitals. The computers were found in the OPD (Out Patient Department) and in the accounts department. It was noted that there were no computers in the doctors' consulting rooms. These computers were rather used by the hospitals administrative staff for capturing patients' demographic information and revenue collection. It was also noted that the Internet connection in some of the hospitals (Taung, Ganyesa, Revilon, Bloemhof) was very slow and often down for two to three times per week.

### 4.2. Doctor's workload

It was noted that the average size of the rural hospitals has a direct link with the number of doctors deployed in the hospital. The average number of doctors of each rural hospital is three doctors per hospital. Each of these doctors serves an average of 18000 patients per month. Ninety percent of these hospitals do not have specialist doctors and most hospital cases have to be referred to the provincial or national hospitals. There is high number of patients whose nature of illness requires specialist attention from these hospitals and this put enormous pressure on rural doctors to consult with their specialist counterparts for professional advice. This high number of patients excludes those who have been admitted to the wards.

### 4.3. Methods of consultation and frequency of consultation among doctors

Consultation doctors from different hospitals are done through face to face methods. This is done through scheduled meetings between these doctors. Consultation is also done through telephonic means, SMS's or e-mails which are computer-mediated. An average of 15 consultations between a doctor and a specialist doctor through face to face or telephone per week was revealed during the interview. Issues regarding doctor to doctor social consultation were done through cell phones and it was frequent.

### 4.4. Task type and speed of transmission of task using CMT

It was noted that doctors in remote hospitals perform numerous tasks which require them to seek professional help using the computer as a mediated tool. These tasks include dental, physiotherapy, occupational therapy, oncology, psychological counseling, x-ray, dietetics and speech therapy. Despite the wide range of services which these hospitals offer, there are no computer-mediated tools like e-consultation, e-referrals, e-prescription, and e-patient record to assist the doctors in their clinical duties. The only computer-mediated tool available is the e-mail. However, transmission of information through e-mails is very slow due to poor Internet connectivity and constant interruption and cut off of electricity supply.

### 4.5. Doctors' satisfaction of using a CMT

All the doctors (n=10) interviewed expressed their dissatisfaction about the use of the computer-mediated tool (e-mail) and cited the slow transmission speed of the Internet and down turn of Internet connectivity. Again the doctors expressed that consultation which involves the sending of patients' images such as x-rays and pathological images for expert (specialist doctor) opinion were difficult because of poor and slow Internet connectivity and lack of software application.

## 5. DISCUSSION

This section discusses the findings from the perspective of each of the characteristics proposed by media synchronization theory (MST) which are rehearsability, reprocessability, parallelism and transmission velocity.

### 5.1. Rehearsability

The ability of doctors to rehearse a message produces an understandable message even if the language skills are less than perfect. Allowing time by the doctors to review spellings of the message improves the structure of the message and makes communication more precise and efficient. It reduces barriers and makes communication through the CMT better. However, the only CMT available to doctors in the rural hospitals is e-mails which is not for supporting clinical activities and therefore requires less rehearsability.

### 5.2. Reprocessability

Reprocessability allows doctors to re-examine the message to be communicated before sending it. It helps doctors to form a better understanding of what is to be communicated, and assists them to review additional information before sending it. The usage of e-mails by doctors in the hospitals gives them enough opportunity to re-examine their messages before sending them. However, doctors used their e-mails as repository for important information which other CMT like e-patient health record system, or e-referral system could have been appropriate for storage of information.

### 5.3. Parallelism

This is how the CMT can integrate work activities simultaneously. The doctors indicated that the ability to perform two or more tasks at the same time using CMT could reduce the enormous pressure emanating from multi tasks. For example, sending e-mails and transmitting patient information simultaneously can provide a high degree of parallelism and can be considered desirable as the CMT performs such multi functions.

### 5.4. Transmission velocity

Transmission velocity refers to the speed at which CMT delivers messages from doctors in remote rural areas to specialist doctors and vice versa. The findings revealed that the speed of Internet connectivity was very slow and sometimes not available due to power outages and poor network infrastructure. The unreliability of bandwidth and the heavy workload of doctors will cause e-consultation in this region not to be supported by real-time communication. The frequency at which consultation is done telephonically and via sms is very high but unfortunately consultations which involve the sending of patients' images such as x-rays and pathological images for specialist doctor's opinion were difficult to transmit because of poor and slow Internet connectivity in these hospitals. The velocity at which data should be generated and transmitted to support real time consultation, e-prescription, e-referrals and e-patient records should be enormous and fast.

## 6. THE NEED FOR HYBRID CLOUD COMPUTING

Based on these findings the researcher proposes synchronous and asynchronous consultation (storage and discussion-centric system) based on cloud computing architecture for the hospitals. IT infrastructure and application will be consumed and shared from the clouds by the various hospitals. A hybrid cloud computing is proposed due to the nature and sensitivity of health data. A hybrid computing is driven by economies of scale in which there is a pool of virtualized and dynamically-scalable computing power, providing storage, platforms and services on demand to the hospitals [14].

Source: [18]

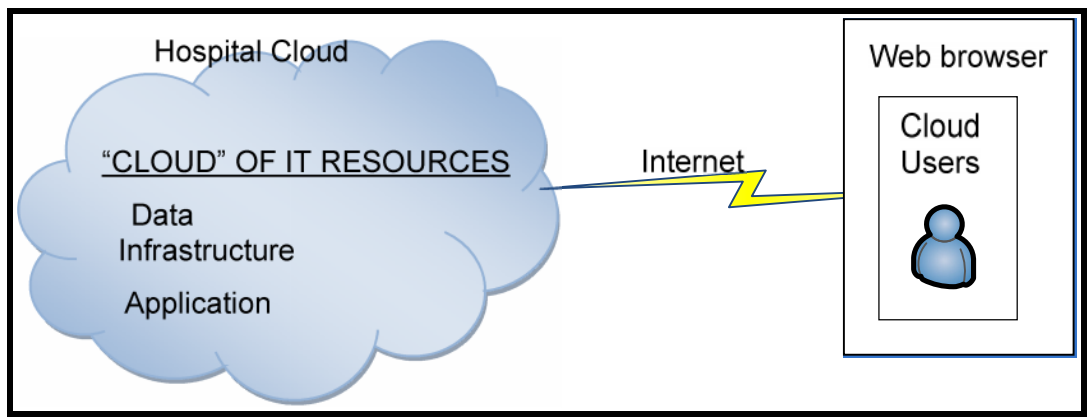

Figure 2: Cloud computing for hospitals

Doctor's communication received through e-mails or discussion can be securely stored in the clouds. Patient information can also be stored in a database in the clouds which will only allow users with proper identification to access it. Cloud infrastructure service also known as "infrastructure as a service (Iaas)" will deliver computer infrastructure to the various hospitals as a service along with raw (block) storage and networking. Hospitals will not purchase servers, software, data centre space, or network equipment but instead buy those resources as a full outsourced service. The cloud application service or software as a service (SaaS) will deliver software services to the hospitals over the Internet, eliminating the need to install and run the application on the doctors' own computers, therefore, simplifying maintenance and support. Such software can include e-mail and Internet base communication, storage and retrieval of data (textual data, voice and video), image acquisition software and software for tele education.

Switching over to cloud computing will give more time to doctors to spend on clinical duties. To switch from traditional health services to cloud-based services, the model proposed by [15] becomes relevant and this model applies the four stages (identification, evaluation, action and follow-up) as indicated in section 2.1. An asynchronous communication system through cloud computing can help doctors share professional knowledge, a doctor located in a remote rural hospital can quickly consult a specialist doctor in an urban area for professional advice as illustrated in Figure 3 below.

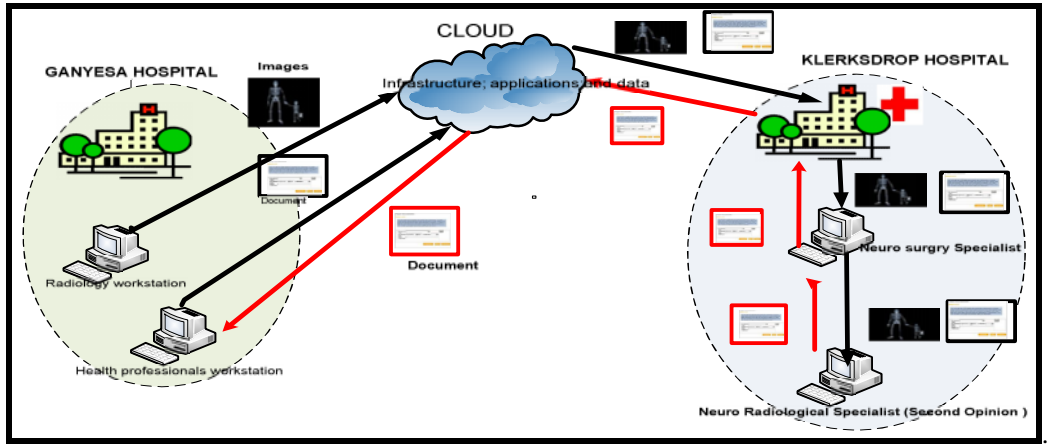

Figure 3: Process of computer-mediated asynchronous communication

Figure 1 shows an example of a doctor in a rural hospital (Ganyesa Hospital) who needs a specialist doctor's advice on a radiology image taken at the hospital. The doctor sends the image or document to the specialist at Klerksdorp Hospital via the file database server stored in the cloud. A neuro-surgery specialist examines it and forwards it to a neuro-radiology specialist for a second opinion. The neuro-radiology specialist gives his opinion to the neuro surgeon who stores the information in the hospital file database server in the cloud and sends a response back to the doctor at Ganyesa Hospital.

## 7. CONCLUSION

Having reviewed the problems of computer-mediated communication in rural hospitals in the North West Province, unpacked the concept of Media Synchronicity Theory, investigated ICT infrastructure, doctors' workload, task types, speed of Internet transmission and doctors' dissatisfaction of not benefiting from computer-mediated communication, it was noted that Computer-mediated asynchronous and synchronous communication between doctors is strongly affected by slow Internet connectivity speed which results in delayed e-mails, slow servers, and constant outages of electricity power supply in rural hospitals. These challenges do not only affect computer-mediated communication but also prevent the implementation of any other e-health services like e-prescription, e-referral and e-patient record keeping. Therefore the following recommendations and conclusion are made:

- Installation of stand by generator for electricity in rural hospitals

- Again because of the pivotal role which computer-mediated communication and other e-health services can play in improving healthcare delivery in rural hospitals, this article recommends computer-mediated communication based on hybrid cloud computing.

- Institute an implementation plan base on [15] health care cloud computing strategic plan model stage 3. At this stage the implementation plan must emphasis the following steps:

    Step 1: Determine the cloud service and deployment model; In this instance the service type (SaaS, or IaaS ) and deployment model (hybrid) cloud are selected.
    Step 2: Compare different cloud provides in terms of pricing schemes, audit procedures, privacy and security.
    Step 3: Obatain assurance from selected cloud provider in terms of quality service
    Step 4: Consider future data migration either to another provider or back to in-house
    Step5: Start a pilot implementation to verify the advantages of hybrid cloud computing for the hospitals.

In conclusion, this article has offered a new approach to computer-mediated asynchronous and synchronous communication for doctors in remote rural hospitals and if embraced will make ICT enhance the delivery of effective healthcare services to the people of North West Province and South Africa as a whole.

**Authors**

Dr Alfred Coleman is a PHD holder in Information Technology and a senior lecturer in school of computing in University of South Africa. His specialization area include e-health an business Informatics

Professor M Heselman is a PHD holder in Information systems and researcher Professor of Living Labs Meraka Institute, CSIR in South Africa.

Mary Coleman is a Masters degree holder and a lecturer at Medical School of South Africa University of Limpopo.